\documentclass[useAMS,usenatbib]{mn2e}
\usepackage{epsfig,amsmath,amssymb,amsfonts,mathrsfs,latexsym,graphicx,lscape}
\usepackage{rotating}
\usepackage{fixltx2e}
\usepackage{longtable}
\usepackage{natbib}
\usepackage{cite}

\newcommand{\xdpsno}{97 } 
\title{On the modulation of low frequency Quasi-Periodic Oscillations in black-hole transients}
\author[Pawar et al.]{\parbox[t]{\textwidth}{\raggedright Devraj D. Pawar
$^1$ \thanks{devrajdp@gmail.com},~ 
Sara Motta$^2$, 
K. Shanthi$^3$, 
Dipankar Bhattacharya$^4$, \\
Tomaso Belloni$^5$
}\\
\vspace{10pt}\\
$^1$ R. J. College, Mumbai, 400086, India\\
$^2$ ESAC, European Space Astronomy Centre, Villanueva de la Ca\~nada, E-28692 Madrid, Spain,\\
$^3$ UGC Academic Staff College, University of Mumbai, Mumbai 400098, India\\
$^4$ Inter University Centre for Astronomy and Astrophysics, Pune 411007, India,\\
$^5$ INAF-Osservatorio Astronomico di Brera, via E. Bianchi 46, 23807 Merate, Italy
}
\begin{document}
\maketitle
\begin{abstract}
We studied the properties of the low-frequency quasi-periodic oscillations detected in a sample of six  black hole candidates (\mbox{XTE~J1550--564}, \mbox{H~1743--322}, \mbox{XTE~J1859+226}, \mbox{4U~1630--47},\mbox{GX~339--4}, \mbox{XTE~J1650-500}) observed by the Rossi XTE satellite. We analyzed the relation between the full width half maximum and the frequency of all the narrow peaks detected in power density spectra where a type-C QPO is observed. Our goal was to understand the nature of the modulation of the signal by comparing the properties of different harmonic peaks in the power density spectrum. 
We find that for the sources in our sample the width of the fundamental and of the first harmonic are compatible with a frequency modulation, while that of the sub-harmonic is independent of frequency, possibly indicating the presence of an additional modulation in amplitude. We compare our results with those obtained earlier from \mbox{GRS~1915+105} and \mbox{XTE~J1550-564}.

\end{abstract}
\section{INTRODUCTION}
\label{sec:intro}
The study of the variable X-ray emission from X-ray binaries has matured very rapidly in the last decades with the launch of several X-ray satellites in the 90's, which led to a clear observational picture of black hole candidates in X-ray binaries (BHBs). In particular a rich phenomenology of fast X-ray variability has been observed. Since their discovery, black hole X-ray binaries (BHB) have been known to 
display variability on various time scales \citep{Elvis75} and low-frequency quasi periodic modulation of the light curves (known as quasi-periodic oscillations, QPO) from these systems were already found in several sources observed by the Japanese satellite \textit{Ginga}.
Sixteen years of operation of the Rossi X-ray Timing Explorer (\textsc{rxte}, December 30, 1995 to January 3, 2012) satellite have contributed immensely to the understanding of these systems and led to an extraordinary progress in the knowledge of the variability properties of BHBs \citep[see e.g.][]{Vanderklis06, Remillard06, Belloni10}.

The typical power density spectrum (PDS) from a BHB is composed by a variable band-limited noise continuum and by a number of QPOs observed both at low (mHz to few tens of Hz) and high (tens to hundreds Hz) frequencies \citep{Vanderklis06}. QPOs are thought to originate in the innermost regions of the accretion flows around stellar-mass black holes and the study of their properties can provide important clues on the physics of accretion onto BHBs and on the behavior of the matter in presence of strong gravitational fields. Various models have been proposed in order to explain the quasi periodic variability and the noise continuum (see \citet[][]{Vanderklis06} for a review and \citet[][]{Ingram11} for a more recent model), but none of them, at the present time, can give a comprehensive explanation of all the observed phenomena.

{In black-hole binaries, low frequency QPOs (LFQPOs) are detected in the frequency range $\sim 0.1 - 30$ Hz; they are classified as type A, B and C QPOs, \citep[see][]{Casella05, Motta11}.}
Here we focus on the type-C LFQPOs; for a complete classification of QPO types in black hole binaries see \citet[][]{Casella05, Motta11, Motta12} and \citet[][for a review]{Belloni11}. 
Type-C QPOs are observed in the hard and hard intermediate state of BHBs in outburst as strong, narrow peaks in the power density spectra with centroid frequency between $\sim$0.1 and $\sim$30 Hz \citep[see][and references therein]{Motta11}. LFQPOs usually appear as a complex structure composed by a primary peak (whose frequency is usually associated to the fundamental frequency of oscillation) and by a number of additional peaks whose frequency is in harmonic relation with that of the that of the primary peak (therefore associated to harmonics and sub-harmonics of the fundamental frequency). These secondary peaks are commonly interpreted as the result of the Fourier decomposition of the non-sinusoidal quasi-periodic signal responsible of the appearance of the LFQPOs. 
However, questions whether there exists a harmonic relation between different QPO peaks and on the identification of the fundamental peak have been raised in recent works \citep[see][]{Rao10, Rodriguez11a, Ratti12}. 

Here we investigate the relation between the properties of the fundamental peak and of the harmonic content component of the type-C LFQPOs detected in a sample of six BHBs observed by \textsc{rxte}: \mbox{XTE~J1550--564}, \mbox{H~1743--322}, \mbox{XTE~J1859+226}, \mbox{4U~1630--47}, \mbox{GX~339--4}, \mbox{XTE~J1650--500}; see reviews by \citet{Remillard06, Belloni11} and \citet{Vanderklis06} for details of these sources. 

\section{Observations}
\label{sec:observations}
From the \textsc{rxte} archive we select a sample of six BHBs where clear type-C QPOs have been reported in the hard and hard-intermediate state. 
We analyzed \xdpsno \textsc{rxte} Proportional Counter Array (\textsc{pca}) 
\citep{Zhang93, Jahoda06} observations of black hole candidates  
\mbox{XTE~J1550--564} \citep{Rao10},
\mbox{H~1743--322} \citep{Homan05b},
\mbox{XTE~J1859+226} \citep{Markwardt01},
\mbox{4U~1630--47} \citep{Tomsick00, Trudolyubov01}, 
\mbox{GX~339--4} \citep{Motta11}
and 
\mbox{XTE~J1650--500} \citep{Homan03}. The references report the detection and analysis of the QPOs in these sources.
Table~\ref{table:obsdet} summarizes the details of the observations that we considered. 
We selected only observations where the sources were either in the low-hard state \citep[LHS, see][]{Belloni11} or in the hard intermediate state (HIMS), where LFQPOs are detected \citep[][]{Casella04, Casella05, Motta11}. Then we selected only those observations whose PDS showed at least two peaks that appeared to be integer multiples of each other. 
We kept only those observations showing type-C QPOs \citep[classified according to][]{Casella04,Motta11,Motta12}; 
type-C QPOs are observed over a large frequency range ($\sim$0.1 - 30 Hz, variable depending on the source), while type-B and -A QPOs only show very small variations in frequency \citep[see][]{Motta11}; type-C QPOs almost always show harmonic peaks and type-B are also usually accompanied by a harmonic, while such harmonics are not usually seen for type-A QPOs.
A sample of type-C QPOs is suitable for our purpose to study the behaviour of 
the multiple QPOs in the same PDS as a function of their frequency as they 
span a relatively larger frequency range compared to other types of LFQPOs. 

{For our work, it is important to identify the peak corresponding 
to the fundamental frequency. Usually, the strongest peak is the 
fundamental, but there can be exceptions. We checked all PDS and derived 
the identification of the fundamental from works in the literature
: \mbox{XTE~J1550-564}
-- \citet{Remillard02}; \mbox{H1743-322} -- \citet{McClintock09}; 
\mbox{XTE~J1859+226} -- \citet{Casella04}; \mbox{4U~1630-47} -- 
\citep{Tomsick05}; \mbox{XTE~J1650--500} -- \citet{Homan03};
\mbox{GX~339--4} -- \citet{Motta12}}.

\begin{figure}
\begin{center}$
\begin{array}{c}
\includegraphics[angle=270.0, width=3.4in]{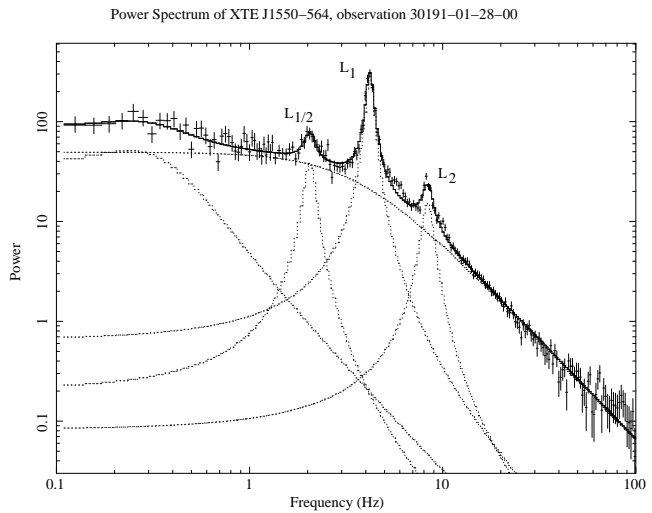} 
\end{array}$
\end{center}
\vskip 0.5cm
\caption{Typical PDS (Observation 30191-01-28-00 of \mbox{XTE~J1550--564}) and model used to fit it. The model is constituted by a broad component centered at zero Hz (sometimes more than one broad component is used to properly describe the noise) and three different narrow Lorentzians, dubbed L$_{1/2}$, L$_1$ and L$_2$, where the frequencies are related according to the following relations: L$_{1/2} \approx 0.5~$L$_1$ and L$_2 \approx 2~$L$_1$ within errors.
}
\label{fig:pds}
\end{figure}

\subsection{Fourier timing analysis and PDS modeling}\label{sec:time}
We used high time resolution mode \textsc{rxte}/\textsc{pca} data for our timing analysis \citep{Zhang93, Jahoda06}. For each observation we chose the data modes in order to have the maximum possible time resolution and the energy range we are interested in. 
We rebinned the light curves to obtain a Nyquist frequency of 2048 Hz and we selected events in the 
channels 0--35; these channels corresspond to energy range $\sim$2--13 keV for the observations of year 1998 to 1999 and $\sim$2--14 keV for later observations in Table~\ref{table:obsdet}, see \citet{Jahoda06} for details of \textsc{rxte}/\textsc{pca}.
We used the \textsc{ghats}\footnote{\textsc{ghats}, available at http://www.brera.inaf.it/utenti/ belloni/GHATS\_Package/Home.html} package under \textsc{gdl} to produce PDS for continuous 32 seconds long data segments  (leading to a minimum frequency of $1/32~s = 0.03125~$Hz) and we averaged them to obtain one single PDS per observation. PDS were normalized according to \citet[][]{Leahy83} and converted to square fractional rms \citep[see][]{Belloni90}. 
{The component due to the Poissonian noise was computed from the deadtime value, the average source count rate and the Very Large Event (VLE) count rate \citep[see][for more details]{Zhang95, Boutloukos06}.}

PDS fitting was carried out with the standard XSPEC fitting
package by using a one-to-one energy-frequency conversion and
a unit response. Following \citet{Belloni02}, we fitted the noise
components with a variable number of broad Lorentzian shapes. The QPOs were fitted with a number of Lorentzians depending on the presence of harmonic peaks. For each Lorentzian component we calculated the characteristic frequency, given by $\nu_{max} =
\sqrt{\nu_{0}^2+(FWHM/2)^2} = \nu_0\sqrt{1+1/(4Q^2)}$
\citep{Belloni02}.  
We fit the PDS to obtain reduced $\chi^2 \approx  1 $ and report 
QPOs with significance $N/\sigma _N \textgreater 3 \sigma$, where N is the normalisation and $\sigma _N$ is its one sigma uncertainty on N obtained with the chi-squared minimization.

In most of the cases, one single Lorentzian is sufficient to fit each peak in the PDS. However, in some cases, more than one Lorentzian component is necessary,  usually when the QPO frequency changes on time scales shorter than the observation time (see \citet{Markwardt99} and references within for details).
For those cases, following \citet{Ratti12}, we report the frequency of the 
Lorentzian with higher rms as the QPO frequency. In these cases, 
the FWHM is estimated as the FWHM of the sum of the two 
Lorentzians and errors on FWHM are calculated as quadratic errors of 
the individual components.
In Figure~\ref{fig:pds} we show a typical PDS (observation 30191-01-28-00 of \mbox{XTE~J1550--564}), fitted with a model of multiple Lorentzians. One or more zero centered Lorentzians are required to fit the low frequency noise. We refer to the Lorentzians used to fit QPOs as
L$_{1/2}$,
L$_{1}$ and
L$_{2}$
respectively, where L$_{1}$ is the primary peak associated to the fundamental frequency, and L$_{1/2}$ and L$_{2}$ are associated to the sub-harmonic and the second harmonic of the fundamental, respectively. The frequencies of the narrow peaks are in the ratio 
\mbox{L$_{1/2}$:L$_{1}$:L$_{2} \simeq 1:2:4$} within the errors and the rms of L$_{1}$ is always greater than L$_{1/2}$ and L$_{2}$.
\section{Results}\label{sec:results}
\begin{figure}
\begin{center}$
\begin{array}{c}
\includegraphics[width=3.4in]{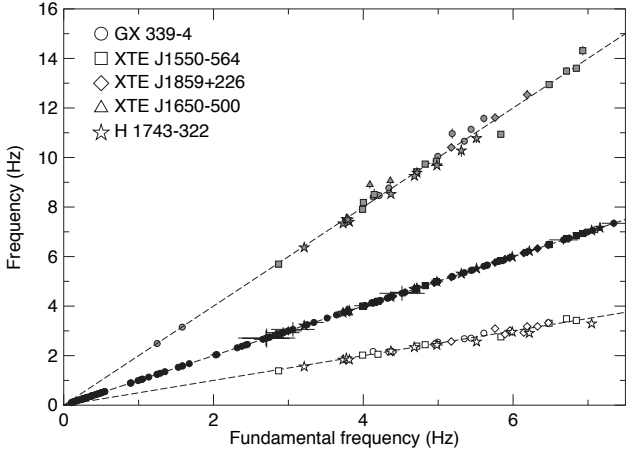}
\end{array}$
\end{center}
\caption{
The harmonic relation of the QPOs detected in the sample of BHBs in Table~\ref{table:obsdet}. QPOs for sources \mbox{GX~339--4}, \mbox{XTE~J1550--564}, \mbox{XTE~J1859+226}, \mbox{H~1743--322}  and \mbox{XTE~J1650--500} are represented as circles, squares, diamonds, triangles and stars respectively. The symbols are filled black, grey and are empty for L$_1$, L$_2$ and L$_{1/2}$ respectively and the dashed lines correspond to slopes 0.5, 1 and 2.
}
\label{fig:qpos}
\end{figure}

In Figure~\ref{fig:qpos} we plot the centroid frequencies of L$_{1/2}$, L$_1$ and L$_2$ QPOs as a function of the centroid frequency of the fundamental (L$_1$) for \mbox{GX~339--4}, \mbox{XTE~J1550--564}, \mbox{XTE~J1859+226}, \mbox{H~1743--322}, \mbox{XTE~J1650--500}, represented as circles, squares, diamonds, empty triangles and stars respectively. 
The symbols are filled black, grey and empty for L$_1$, L$_2$ and L$_{1/2}$ respectively.
The dashed lines correspond to slopes 0.5, 1 and 2; L$_{1/2}$ and L$_2$ QPOs lie around lines with slopes 0.5 and 2 respectively, with a small but significant scatter.
We detected L$_{1/2}$ or L$_2$ or both simultaneously with L$_1$ in the power spectra of 69 observations (\#1 to 72, except \#37, 59, 60) of the total \xdpsno observations in our sample (see Table~\ref{table:obsdet} for details).

\begin{figure}
\begin{center}$
\begin{array}{c}
\includegraphics[width=3.4in]{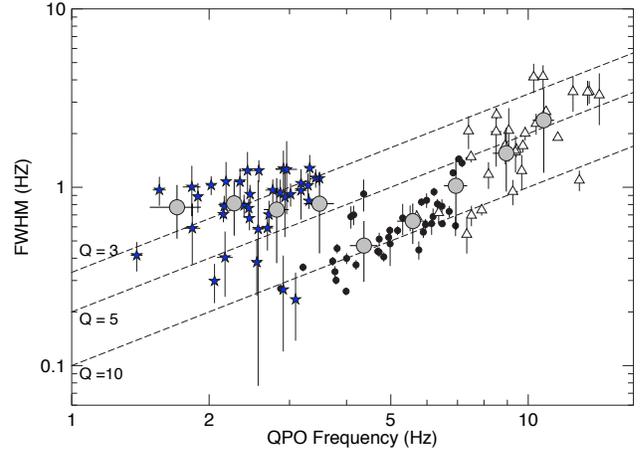} \\
\includegraphics[width=3.4in]{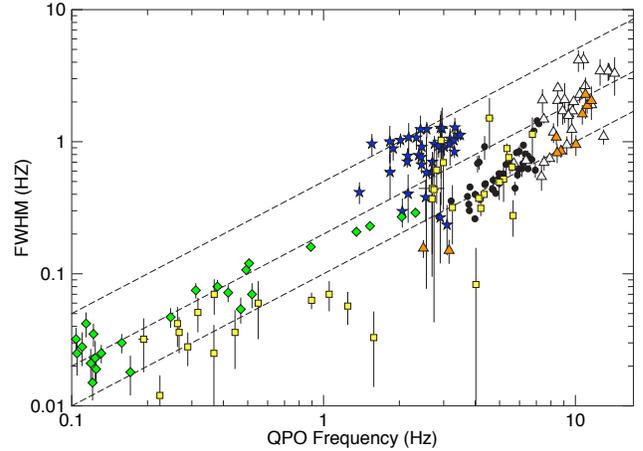}
\end{array}$
\end{center}
\caption{
\textit{Top panel :} The FWHM as a function of centroid frequency of type-C  QPOs detected in the power spectra of black holes in our sample such that, either L$_{1/2}$ or L$_2$, or both L$_{1/2}$ and L$_2$ are simultaneously detected with L$_1$ (observations 1 to 72 in Table~\ref{table:obsdet}). Stars, filled circles, empty triangles respectively represent L$_{1/2}$, L$_1$, L$_2$ QPOs. The dashed lines correspond to constant Q. The large grey filled circles are the result of a rebinning of all points into broad logarithmic bins in frequency.
\mbox{\textit{Bottom panel :}} Stars, filled circles, triangles respectively represent L$_{1/2}$, L$_1$, L$_2$ such that, either L$_{1/2}$ or L$_2$, or both are simultaneously detected with L$_1$. Diamonds represent type-C QPOs detected in \mbox{GX~339--4}, where no harmonics were detected \citep{Motta11} and squares, filled triangles respectively represent the fundamental, harmonic of the type-C QPOs detected in \mbox{GRO~J1655--40}, \citep{Motta12}. In both panels the dashed lines correspond to lines of constant Q = 3, 5, 10.
}
\label{fig:fwhmq}
\end{figure}

In Figure~\ref{fig:fwhmq} we show the  FWHM of the fundamental peak and its harmonics plotted as a function of the centroid frequency of each peak. 
In both panels we overplot dashed lines corresponding to Q = 3, 5 and 10, where Q is the \textit{quality factor} and it is defined as $Q = \nu_0/FWHM$. It gives a measure of the coherence of a peak.  
The symbols do not distinguish between the sources. This is done with the purpose to make the general trend of the points clearer. 
In the top panel of Fig. \ref{fig:fwhmq} we plot the points from observations, that have at least L$_{1/2}$ or L$_2$ in addition to L$_1$ in the same PDS (Table~\ref{table:obsdet}). L$_{1/2}$, L$_1$ and L$_2$ peaks are represented as stars, filled circles and empty traingles respectively. The frequency range covered is $\sim$2--16 Hz. The large grey filled circles are the result of a rebinning of all points into broad logarithmic bins in frequency. Each point has a frequency equal to the mean of all the points included in a single bin, while the error bars represent the scatter as the square root of the variance in frequency. 
In the lower panel of Figure~\ref{fig:fwhmq} we show the same data of the upper panel, but we also include the remaining QPOs of \mbox{GX~339--4} \citep{Motta11}, (73 to 98 in Table~\ref{table:obsdet}) and QPOs detected in
\mbox{GRO~J1655--40} reported in \citet{Motta12}. The additional points from \mbox{GX~339--4} (filled diamonds) extend the frequency down to 0.1 Hz. Since they occur as single peaks, i.e. without second, third and sub-harmonics, they cannot  be classifed as L$_{1/2}$, L$_1$ or L$_2$. The \mbox{GRO~J1655--40} peaks are detected in the frequency range $\sim$0.2 to 15 Hz and the L$_1$ and L$_2$ components are respectively represented as filled squares and filled triangles. 

\begin{table}
\centering
\caption{
Parameters to test whether the frequency-FWHM relation is linear or constant.
L$_1$ + L$_2$ is the data of fundamental and harmonic considered together since they 
seem to have the same nature in the frequency-FWHM plot (Figure~\ref{fig:fwhmq}).
 }
\begin{tabular}{l c c c c}
\hline
\hline 
QPO & \multicolumn{2}{c}{$\chi^2$}         & dof & Pearson's coefficient \\
    & linear            &     constant     &     & \\
[0.5ex]
\hline
L$_1$         &  1933 & 4106  & 85  & 0.61 \\
L$_2$         &  706  & 2051  & 63  & 0.55 \\
L$_{1/2}$     &  258  & 338   & 52  & 0.44 \\
L$_1$ + L$_2$ &  4830 & 23660 & 148 & 0.78 \\
[1ex]
\hline
\end{tabular}
\label{table:nonlin}
\end{table}

{
From the frequency--FWHM correlation in Figure~\ref{fig:fwhmq} one sees that, although there is significant scatter, the fundamental follows an upward trend which is extended by the harmonics points and is steeper than the lines of constant Q, indicating lower Q at higher QPO frequencies, while the subharmonic points do not show a visible trend. We have performed constant and linear fits to the data, but the large scatter prevents a statistical analysis of the relative goodness of the models.
We investigated whether the frequency--FWHM correlation is linear ($FWHM = A\cdot\nu_0 + B$) or constant for L$_{1/2}$, L$_1$, L$_2$.
Since the L$_1$ and L$_2$ seem to have the same frequency--FWHM correleation (Figure~\ref{fig:fwhmq}), we also investigate the linear and constant fit for both together (L$_1$ + L$_2$). 
The $\chi^2$ for the fits indicate that the linear fit is more appropriate for L$_1$, L$_2$, (L$_1$ + L$_2$) and the constant for L$_{1/2}$. However the scatter renders the fits statistically not-significant (see Table~\ref{table:nonlin} for details). 
}
\section{DISCUSSION}\label{sec:discussion}
We analyzed \xdpsno archival \textsc{rxte} observations of six BHBs 
in the hard and hard-intermediate state and studied the FWHM--frequency relation of the type-C QPOs detected in these sources. 
In 69 observations we detect more than one peak superimposed to a broad-band flat-top noise, (see Table~\ref{table:obsdet} for details of the QPOs). The secondary peaks L$_{1/2}$, L$_2$ have centroid frequencies consistent with $\sim$0.5, $\sim$2 times the frequency of the fundamental L$_1$. 
In Figure~\ref{fig:qpos} we see more clearly that L$_{1/2}$, L$_1$ and L$_2$ have frequencies in the ratio 0.5:1:2, therefore they form an harmonic series in frequency.
{The presence of harmonically related peaks in a PDS might be the result of the Fourier decomposition of the signal. However, a scenario where such peaks are independent signals, cannot be ruled out.} 

From the top panel of Figure~\ref{fig:fwhmq} we observe that the FWHM of L$_1$ and L$_2$ increases with frequency of the QPO. The FWHM of L$_{1/2}$ however has a weak dependence on frequency. These trends are seen more clearly after rebinning the points in frequency (large gray points in Figure~\ref{fig:fwhmq}). The resulting large grey points indicate a monotonically increasing trend of the FWHM with frequency for L$_1$, L$_2$ and also that the FWHM of L$_{1/2}$ QPOs has a very weak or no dependence on frequency. 
The QPOs in this plot cover the frequency range $\sim$2--16 Hz. In order to test whether the same relation extends to lower frequencies, we include type-C QPOs \citep[from][]{Motta11, Motta12} which extend frequency down to $\sim$0.1 Hz, (bottom panel in Figure~\ref{fig:fwhmq}).
Adding points for L$_1$ and L$_2$ at lower QPO frequencies from \citet{Motta11, Motta12} we show that the FWHM-frequency relation for these two components extends down to $\sim$0.1 Hz.
The lower frequency QPOs ($\sim$0.1 Hz) correspond to harder energy spectra, when the noise level is high and the QPO is not as well resolved from the continuum as in the softer states. In such cases the QPO parameters are less constrained, (especially the FWHM) and this may be the cause of the scatter observed in the FWHM of QPOs with frequency around $\sim$0.1 Hz.

The coherence of a QPO is quantified by the quality factor Q. The width of a quasi-periodic signal can be ascribed to either amplitude modulation (AM) or frequency modulation (FM) or a combination of the two. 
{The often-considered case of finite time of the oscillation is a special case of amplitude modulation.
}
If only a single peak is seen in the PDS, it is not possible to establish whether the broadening of the peak is due to amplitude or frequency modulation. However if a harmonic is also present, then a constant FWHM of the peaks indicates amplitude modulation and a constant Q indicates frequency modulation. We observe that L$_1$ and L$_2$ are almost along the line of constant Q (top panel of Figure~\ref{fig:fwhmq}). If they were exactly along the constant Q line, we could conclude that they are frequency modulated. However they are broader at higher frequencies. 
{One possibility is that this is due to an additional amplitude modulation, although different combinations of FM and AM cannot be excluded.
}
The sub-harmonics (L$_{1/2}$) have a constant FWHM and are broader, indicating that the sub-harmonic must be subject to an additional amplitude modulation, 
{as the FWHM of these points show a large scatter, it is however still possible that an underlying frequency modulation is present.}

A similar approach to the one we follow in this paper was presented by \citet{Rao10} and \citet{Ratti12} for the type-C QPOs detected in \mbox{XTE~J1550--564} and \mbox{GRS~1915+105} respectively. They discuss the FWHM frequency relation between peaks, flat topped noise and peaked noise detected in the PDS. In our paper, we consider only the L$_{1/2}$, L$_1$, L$_2$ peaks detected in the sample of BHBs listed in Table~\ref{table:obsdet}.
\citet{Rao10} in their study of QPOs detected in the BHB \mbox{XTE~J1550--564}, find that the four features L$_{ft}$, L$_s$, L$_{F}$, L$_{pn}$, L$_{h}$ form a harmonic series in frequency (where L$_s$, L$_F$, L$_h$ correspond to L$_{1/2}$, L$_1$, L$_2$ respectively in this paper; L$_{ft}$ is the flat topped noise and L$_{pn}$ is the peaked noise component, see Figure 1 in \citet{Rao10} for details and \citet{Belloni02}, \citet{Vanderklis06} for terminology and classification of features detected in PDS of BHBs). \citet{Rao10} conclude that not all the features in the harmonic series follow the same FWHM frequency relation. L$_1$ and L$_2$ have same evolution in FWHM with frequency, where as L$_{1/2}$ does not follow the same trend. Based on this difference and other analysis of the rms, energy spectra in frequency, they suggest that L$_{1/2}$ is not an harmonic artefact, but is an oscillation which has additional modulations than that of L$_1$ and L$_2$. Our result agrees with this interpretation.

In the case of LFQPOs in \mbox{GRS~1915+105}, \citet[][]{Ratti12} conclude that for L$_1$ and L$_2$, the coherence decreases as frequency increases, which suggests frequency modulation of these components. While L$_1$ and L$_2$ are indistinguishable in their dependence of coherence on frequency, L$_{1/2}$ has a much lower coherence than it would, had it followed the same coherence-frequency relation. The constant width of L$_{1/2}$ indicates amplitude modulation, which is large enough to overwhelm any possible frequency dependence. We have a similar conclusion for QPOs across our sample of BHBs, see top panel of Figure~\ref{fig:fwhmq}. From the lower panel of Figure~\ref{fig:fwhmq}, it is clear that only the L$_{1/2}$ components are amplitude modulated, while the coherence of the fundamental and its harmonic is frequency dependent even at lower frequencies.

Comparing our results with \citet{Rao10} and \citet{Ratti12} we find that the FWHM-frequency relations of type-C QPOs reported by them are also seen across the sample of sources which we have analysed. The sub-harmonic and second harmonic components are amplitude modulated, which reduces their quality factor much more than the fundamental one, which is frequency modulated. Moreover the sub-harmonic has a much larger amplitude modulation compared to the other QPOs in the harmonic series and this effect is seen across a sample of black hole binaries.
{Over the dependence of FWHM and frequency of L$_{1/2}$, L$_1$ and L$_2$, we detect a scatter in the FWHM. This scatter is not related to any observational effects such as observation duration as we detect no correlation in the FWHM and observation duration in case of QPOs which are nearly equal in frequency. The investigation of the scatter 
{which can have a significant impact on the interpretation of the data} is beyond the scope of this paper and will be done in the future on the basis of a larger database of QPOs.}

Whatever the nature of the observed frequencies and the physical origin of the variability, the modulation which gives rise to the width of the peaks detected in the PDS appears to be complex. Both amplitude and frequency modulation seem to be at work on separate harmonics of the signal, yielding a complex pattern which can be tested against theoretical models.

\section{ACKNOWLEDGMENTS}
DP and KS thank IUCAA, Pune, India and INAF--OAB, Merate, Italia
for their support and thank Ranjeev Misra and Gulab Dewangan for
discussions during their visits to IUCAA. 
SM thanks the ESA fellowship program.
TB acknowledges support from INAF PRIN 2012-6. \\

\noindent This research has made use of the General High-energy
Aperiodic Timing Software (GHATS) package developed
by T.M. Belloni at INAF - Osservatorio Astronomico di
Brera. \\

\noindent 
This project is supported by the Department of Science and 
Technology (DST), India, and Ministry of External Affairs, 
Italy under the Joint Indo-Italian project, grant order number
INT/ItalyP-3/2012 (ER). \\

\noindent This research has made use of
NASA's Astrophysics Data System. This research has made use of data
obtained through the High Energy Astrophysics Science Archive Research
Centre Online Service, provided by the NASA/Goddard Space Flight
Center. \\

\noindent 
{We thank the anonymous referee for his or her comments and suggestions.} 
\bibliographystyle{aa}

%
%
\onecolumn
\begin{center}
\begin{longtable}{ll c c c c c c c}            
\caption{
Details of observations used in our analysis.
These observations were obtained over a number of years in different campaigns 
of the Rossi X-ray Timing Explorer mission. 
We report frequency ($\nu_1$) and full width at half maximum ($\Delta$) for the 
Fundamental (L$_1$), harmonic (L$_2$) and sub-harmonic (L$_{1/2}$).
The errors are quadratic means of the $1\sigma$ errors obtained in XSPEC.
A feature which is not significantly detected is indicated by $"--"$.} \\
\hline
&Observation	&Date		&\multicolumn{2}{c}{Fundamental}&\multicolumn{2}{c}{Harmonic}   &\multicolumn{2}{c}{Subharmonic} \\
&		&\tiny{(yyyy-mm-dd)}&$	\nu_0		$&$		\Delta		$&$		\nu_0		$&$		\Delta		$&$		\nu_0		$&$		\Delta		$	\\
&      		&		&	(Hz)		&		(Hz)		&		(Hz)		&		(Hz)		&		(Hz)		&		(Hz)			\\
\hline
\endfirsthead
\caption{continued.}   \\
\hline
&Observation	&Date		&\multicolumn{2}{c}{Fundamental}&\multicolumn{2}{c}{Harmonic}   &\multicolumn{2}{c}{Subharmonic}\\
&		&\tiny{(yyyy-mm-dd)}&$	\nu_0		$&$		\Delta		$&$		\nu_0		$&$		\Delta		$&$		\nu_0		$&$		\Delta		$	\\
&      		&		&	(Hz)		&		(Hz)		&		(Hz)		&		(Hz)		&		(Hz)		&		(Hz)			\\
\hline
\endhead
\hline \multicolumn{8}{r}{{Continued on next page}} \\
\endfoot
\hline
\endlastfoot
     \hline
     \multicolumn{9}{l}{\mbox{XTE~J1550--564}}\\
     \hline
1  &30188-06-11-00	&1998-09-16&$	3.99	\pm	0.01	$&$	0.26	\pm	0.01	$&$	7.91	\pm	0.02	$&$	0.75	\pm	0.05	$&$	2.02	\pm	0.05	$&$	1.03	\pm	0.12	$	\\
2  &30191-01-01-00	&1998-09-18&$	5.84 	\pm	0.02	$&$	0.83	\pm	0.04	$&$	10.94	\pm	0.09	$&$	2.66	\pm	0.20	$&$	2.76	\pm	0.10	$&$	0.96	\pm	0.14	$	\\
3  &30191-01-03-00	&1998-09-20&$	6.88 	\pm	0.03	$&$	1.42	\pm 	0.05	$&$		 --		$&$		 --		$&$	3.84	\pm	0.08	$&$	1.27	\pm	0.22	$	\\
4  &30191-01-06-00	&1998-09-21&$	7.35 	\pm	0.09    $&$	1.34	\pm	0.16	$&$		 --		$&$		 --		$&$	3.26	\pm 	0.11	$&$	2.86	\pm	0.21	$	\\
5  &30191-01-07-00	&1998-09-21&$	5.76	\pm	0.01	$&$	0.64	\pm	0.03	$&$	11.16	\pm	0.09	$&$	2.12	\pm	0.30	$&$	2.59	\pm	0.08	$&$	1.17	\pm	0.22	$	\\
6  &30191-01-15-00	&1998-09-29&$	4.07	\pm	0.01	$&$	0.32	\pm	0.03	$&$	8.14	\pm	0.01	$&$	0.87	\pm	0.06	$&$	2.06	\pm	0.04	$&$	0.78	\pm	0.11	$	\\
7  &30191-01-16-00	&1998-09-29&$	3.16	\pm	0.03	$&$	0.54	\pm	0.04	$&$     5.98	\pm	0.03	$&$	1.75	\pm	0.18	$&$	1.53	\pm	0.03	$&$	0.58	\pm	0.12	$	\\
8  &30191-01-16-01	&1998-09-29&$	2.87	\pm	0.01	$&$	0.27	\pm	0.02	$&$	5.70	\pm	0.02	$&$	0.68	\pm	0.06	$&$	1.39	\pm	0.04	$&$	0.41	\pm	0.08	$	\\
9  &30191-01-27-00	&1998-10-10&$	5.79 	\pm	0.05	$&$	0.32	\pm	0.07	$&$	9.96	\pm	0.06	$&$	0.57	\pm	0.23	$&$	2.68	\pm	0.03	$&$	1.04	\pm	0.14	$	\\
10 &30191-01-27-01	&1998-10-09&$	4.47	\pm	0.01	$&$	0.31	\pm	0.03	$&$	8.97	\pm	0.04	$&$	1.48	\pm	0.24	$&$	2.24	\pm	0.04	$&$	0.63	\pm	0.12	$	\\
11 &30191-01-28-00	&1998-10-11&$	4.20	\pm	0.01	$&$	0.37	\pm	0.02	$&$	8.39	\pm	0.04	$&$	1.24	\pm	0.18	$&$	2.06	\pm	0.02	$&$	0.30	\pm	0.07	$	\\
12 &30191-01-28-01	&1998-10-11&$	4.71	\pm	0.01	$&$	0.43	\pm	0.03	$&$	9.42	\pm	0.04	$&$	1.62	\pm	0.15	$&$	2.39	\pm	0.03	$&$	0.79	\pm	0.12	$	\\
13 &30191-01-28-02	&1998-10-12&$	4.83	\pm	0.01	$&$	0.41	\pm	0.02	$&$	9.93	\pm	0.06	$&$	1.98	\pm	0.25	$&$	2.40	\pm	0.03	$&$	0.71	\pm	0.11	$	\\
14 &30191-01-29-00	&1998-10-13&$	4.98	\pm	0.01	$&$	0.57	\pm	0.03	$&$	9.74	\pm	0.04	$&$	1.71	\pm	0.14	$&$	2.44	\pm	0.03	$&$	0.76	\pm	0.10	$	\\
15 &30191-01-29-01	&1998-10-13&$	6.48	\pm	0.01	$&$	0.62	\pm	0.02	$&$	9.85	\pm	0.05	$&$	2.01	\pm	0.18	$&$	2.45	\pm	0.03	$&$	0.67	\pm	0.10	$	\\
16 &30191-01-30-00	&1998-10-14&$	6.85	\pm	0.02	$&$	1.21	\pm	0.06	$&$	12.94	\pm	0.04	$&$	1.10	\pm	0.14	$&$	3.31	\pm	0.02	$&$	0.84	\pm	0.08	$	\\
17 &30191-01-31-00	&1998-10-15&$	6.71	\pm	0.02	$&$	0.73	\pm	0.05	$&$	13.60	\pm	0.14	$&$	3.43	\pm	0.50	$&$	3.42	\pm	0.04	$&$	1.13	\pm	0.14	$	\\
18 &30191-01-31-01	&1998-10-15&$	6.71	\pm	0.01	$&$	0.73	\pm	0.06	$&$	13.49	\pm	0.16	$&$	3.42	\pm	0.54	$&$	3.49	\pm	0.04	$&$	1.12	\pm	0.10	$	\\
19 &40401-01-57-00	&1999-03-13&$	6.73	\pm	0.04	$&$	1.11	\pm	0.08	$&$	13.54	\pm	0.18	$&$	3.73	\pm	0.40	$&$	3.40	\pm	0.04	$&$	1.11	\pm	0.08	$	\\
20 &50134-01-04-00	&2000-05-12&$	6.93	\pm	0.03	$&$	0.61	\pm	0.08	$&$	14.31	\pm	0.19	$&$	3.29	\pm	1.06	$&$		--		$&$		 --		$	\\
21 &50134-01-05-00	&2000-05-14&$	4.46	\pm	0.02	$&$	0.62	\pm	0.06	$&$	9.21	\pm	0.15	$&$	2.65	\pm	0.66	$&$		--		$&$	  	 --		$	\\
22 &50134-02-01-01	&2000-04-26&$	4.04	\pm	0.01	$&$	0.40	\pm	0.03	$&$	8.19	\pm	0.06	$&$	1.18	\pm	0.20	$&$		--		$&$		 --		$	\\
23 &50135-01-01-00	&2000-05-16&$	4.15	\pm	0.03	$&$	0.70	\pm	0.09	$&$	8.51	\pm	0.24	$&$	2.05	\pm	0.74	$&$		--		$&$		 --		$	\\
     \hline
    \multicolumn{9}{l}{\mbox{XTE~J1650--500}}\\
     \hline
24 &60113-01-12-00	&2001-09-18&$	4.09	\pm	0.03	$&$	0.68	\pm	0.10	$&$	8.91	\pm	0.16	$&$	1.71	\pm	0.37	$&$		--		$&$		--		$	\\
25 &60113-01-12-01	&2001-09-18&$	4.36	\pm	0.03	$&$	0.92	\pm	0.19	$&$	9.07	\pm	0.16	$&$	2.08	\pm	0.69	$&$		--		$&$		--		$	\\
26 &60113-01-12-02	&2001-09-18&$   4.84	\pm	0.03	$&$	0.81	\pm	0.03	$&$	9.77	\pm	1.63	$&$	1.09	\pm	0.14	$&$		--		$&$		--		$	\\
27 &60113-01-12-03	&2001-09-18&$	4.63	\pm	0.06	$&$	1.48	\pm	0.40	$&$	9.03	\pm	0.71	$&$	7.95	\pm	1.75	$&$		--		$&$		--		$	\\
28 &60113-01-12-04	&2001-09-18&$	5.17	\pm	0.05	$&$	0.91	\pm	0.25	$&$	10.51	\pm	0.13	$&$	1.96	\pm	0.43	$&$		--		$&$		--		$	\\
29 &60113-01-13-00	&2001-09-19&$	5.90	\pm	0.26	$&$	0.90	\pm	0.31	$&$	11.53	\pm	0.16	$&$	1.41	\pm	1.05	$&$		--		$&$		--		$	\\
30 &60113-01-13-01	&2001-09-19&$	6.34	\pm	0.06	$&$	1.67	\pm	0.28	$&$	12.20	\pm	0.08	$&$	0.28	\pm	0.38	$&$		--		$&$		--		$	\\
31  &60113-01-13-02	&2001-09-19&$	6.84	\pm	0.08	$&$	1.52	\pm	0.50	$&$	14.21	\pm	0.25	$&$	3.28	\pm	0.93	$&$		--		$&$		--		$	\\
    \hline
   \multicolumn{8}{l}{\mbox{H~1743--322}}\\
    \hline
32 &80138-01-04-00G	&2003-04-06&$	3.73	\pm	0.01	$&$	0.39	\pm	0.02	$&$	7.34	\pm	0.03	$&$	0.54	\pm	0.12	$&$	1.83	\pm	0.08	$&$	1.06	\pm	0.32	$	\\
33 &80138-01-06-00	&2003-04-10&$	3.21	\pm	0.01	$&$	0.36	\pm	0.02	$&$	6.37	\pm	0.03	$&$	0.72	\pm	0.10	$&$	1.56	\pm	0.05	$&$	0.97	\pm	0.18	$	\\
34 &80138-01-07-00	&2003-04-12&$	7.16	\pm	0.02	$&$	1.37	\pm	0.07	$&$		--        	$&$		 --		$&$		--		$&$		--		$	\\
35 &80146-01-02-00	&2003-04-15&$	5.50	\pm	0.02	$&$	0.68	\pm	0.06	$&$     10.80	\pm	0.17	$&$	2.22	\pm	0.35	$&$		--		$&$		--		$	\\
36 &80146-01-03-00	&2003-04-17&$	4.72	\pm	0.01	$&$	0.51	\pm	0.04	$&$	9.38	\pm	0.12	$&$	1.58	\pm	0.37	$&$		--		$&$		--		$	\\
37 &80146-01-03-01	&2003-04-18&$	7.18	\pm	0.03	$&$	1.39	\pm	0.03	$&$		--		$&$		--		$&$		--		$&$		--		$	\\
38 &80146-01-29-00	&2003-05-07&$	5.51	\pm	0.02	$&$	0.68	\pm	0.11	$&$	10.78	\pm	0.19	$&$	4.17	\pm	0.66	$&$	2.56	\pm	0.07	$&$	0.58	\pm	0.51	$	\\
39 &80146-01-30-00	&2003-05-08&$	4.37	\pm	0.01	$&$	0.47	\pm	0.03	$&$	8.52	\pm	0.04	$&$	2.55	\pm	0.14	$&$	2.18	\pm	0.04	$&$	1.08	\pm	0.31	$	\\
40 &80146-01-31-00	&2003-05-09&$	5.31	\pm	0.04	$&$	0.67	\pm	0.13	$&$	10.28	\pm	0.25	$&$	4.15	\pm	0.78	$&$		--		$&$		--		$	\\
41 &80146-01-32-00	&2003-05-10&$	4.98	\pm	0.08	$&$	0.48	\pm	0.12	$&$	9.68	\pm	0.07	$&$	1.24	\pm	0.28	$&$	2.43	\pm	0.10	$&$	1.24	\pm	0.34	$	\\
42 &80146-01-33-00	&2003-05-11&$	6.22	\pm	0.02	$&$	0.94	\pm	0.05	$&$		--		$&$		--		$&$	2.91	\pm	0.12	$&$	1.27	\pm	0.33	$	\\
43 &80146-01-33-01	&2003-05-11&$	5.99	\pm	0.02	$&$	0.85	\pm	0.08	$&$		--		$&$		--		$&$	2.96	\pm	0.16	$&$	1.26	\pm	0.55	$	\\
44 &80146-01-43-00	&2003-05-20&$	3.77	\pm	0.01	$&$	0.34	\pm	0.03	$&$	7.51	\pm	0.03	$&$	0.70	\pm	0.09	$&$	1.89	\pm	0.04	$&$	0.89	\pm	0.17	$	\\
45 &80146-01-43-01	&2003-05-20&$	3.81	\pm	0.01	$&$	0.45	\pm	0.03	$&$	7.40	\pm	0.09	$&$	2.07	\pm	0.43	$&$	1.84	\pm	0.08	$&$	0.59	\pm	0.22	$	\\
46 &80146-01-44-00	&2003-05-21&$	3.79	\pm	0.01	$&$	0.30	\pm	0.02	$&$	7.49	\pm	0.04	$&$	1.48	\pm	0.11	$&$	1.90	\pm	0.04	$&$	0.68	\pm	0.09	$	\\
47 &80146-01-46-00	&2003-05-23&$	4.68	\pm	0.01	$&$	0.44	\pm	0.02	$&$	9.25	\pm	0.05	$&$	0.95	\pm	0.15	$&$	2.34	\pm	0.06	$&$	1.07	\pm	0.21	$	\\
48 &80146-01-47-00	&2003-05-24&$	7.05	\pm	0.02	$&$	1.44	\pm	0.05	$&$		--		$&$		--		$&$	3.29	\pm	0.07	$&$	1.03	\pm	0.17	$	\\
    \hline
   \multicolumn{8}{l}{\mbox{XTE~J1859+226}}\\
   \hline
49 &40124-01-08-00	&1999-10-14&$	4.39	\pm	0.01	$&$	0.48	\pm	0.02	$&$     8.80	\pm	0.04	$&$	1.94	\pm	0.11	$&$	2.15	\pm	0.04	$&$	0.71	\pm	0.07	$	\\
50 &40124-01-09-00	&1999-10-14&$	4.95	\pm	0.01	$&$	0.52	\pm	0.03	$&$	9.87	\pm	0.05	$&$	0.86	\pm	0.19	$&$	2.46	\pm	0.04	$&$	0.92	\pm	0.12	$	\\
51 &40124-01-10-00	&1999-10-14&$	5.76	\pm	0.02	$&$	0.45	\pm	0.05	$&$	11.61	\pm	0.09	$&$	1.90	\pm	0.01	$&$	3.09	\pm	0.05	$&$	0.24	\pm	0.10	$	\\
52 &40124-01-11-00	&1999-10-15&$	5.96	\pm	0.02	$&$	0.62	\pm	0.05	$&$	12.07	\pm	0.10	$&$	1.36	\pm	0.33	$&$	3.02	\pm	0.06	$&$	0.91	\pm	0.14	$	\\
53 &40124-01-15-00	&1999-10-22&$	6.14	\pm	0.03	$&$	0.63	\pm	0.09	$&$	12.21	\pm	0.12	$&$	1.11	\pm	0.46	$&$	2.93	\pm	0.08	$&$	0.88	\pm	0.36	$	\\
54 &40124-01-18-00	&1999-10-20&$	5.89	\pm	0.01	$&$	0.56	\pm	0.04	$&$	11.82	\pm	0.12	$&$	1.54	\pm	0.40	$&$	2.86	\pm	0.05	$&$	0.93	\pm	0.17	$	\\
55 &40124-01-19-00	&1999-10-20&$	5.18	\pm	0.01	$&$	0.57	\pm	0.03	$&$	10.41	\pm	0.07	$&$	2.28	\pm	0.31	$&$	2.57	\pm	0.05	$&$	1.24	\pm	0.18	$	\\
56 &40124-01-20-00	&1999-10-20&$	6.47	\pm	0.02	$&$	0.78	\pm	0.07	$&$	13.02	\pm	0.20	$&$	2.59	\pm	0.75	$&$	3.32	\pm	0.08	$&$	1.28	\pm	0.24	$	\\
57 &40124-01-21-00	&1999-10-21&$	6.33	\pm	0.02	$&$	0.81	\pm	0.05	$&$	12.98	\pm	0.14	$&$	2.28	\pm	0.57	$&$	3.18	\pm	0.05	$&$	1.05	\pm	0.15	$	\\
58 &40124-01-25-00	&1999-10-23&$	6.19	\pm	0.02	$&$	0.68	\pm	0.08	$&$	12.54	\pm	0.19	$&$	3.43	\pm	0.78	$&$	3.18	\pm	0.10	$&$	0.96	\pm	0.30	$	\\
   \hline
   \multicolumn{8}{l}{\mbox{4U~1630-47}}\\
   \hline
59 &80117-01-21-01	&2004-03-04&$	5.11	\pm	0.03	$&$	0.27	\pm	0.08	$&$		--		$&$		--		$&$		--		$&$		--		$	\\
60 &80117-01-22-00	&2004-03-05&$	4.02	\pm	0.02	$&$	0.17	\pm	0.04	$&$		--		$&$		--		$&$		--		$&$		--		$	\\
   \hline
   \multicolumn{8}{l}{\mbox{GX~339-4}}\\
   \hline
61 &60705-01-70-00	&2004-08-13&$	4.10	\pm	0.18	$&$	0.31	\pm	0.07	$&$	8.86	\pm	0.37	$&$	1.02	\pm	0.39	$&$	2.40	\pm	0.11	$&$	0.66	\pm	0.22	$	\\
62 &70109-04-01-00	&2002-05-11&$	5.44	\pm	0.01	$&$	0.76	\pm	0.04	$&$	11.14	\pm	0.07	$&$	1.88	\pm	0.25	$&$	2.71	\pm	0.04	$&$	0.70	\pm	0.14	$	\\
63 &70109-04-01-01	&2002-05-11&$	5.39	\pm	0.01	$&$	0.70	\pm	0.05	$&$	10.90	\pm	0.04	$&$	2.64	\pm	0.55	$&$	2.63	\pm	0.03	$&$	0.98	\pm	0.13	$	\\
64 &70109-04-01-02	&2002-05-12&$	5.35	\pm	0.03	$&$	0.89	\pm	0.08	$&$	10.66	\pm	0.10	$&$	1.62	\pm	0.27	$&$	2.68	\pm	0.08	$&$	0.59	\pm	0.15	$	\\
65 &70110-01-11-00	&2002-05-12&$	5.75	\pm	0.04	$&$	0.65	\pm	0.21	$&$	11.69	\pm	0.20	$&$	2.99	\pm	1.01	$&$	3.03	\pm	0.11	$&$	1.30	\pm	0.50	$	\\
66 &90110-02-01-01	&2004-08-15&$	5.09	\pm	0.06	$&$	0.97	\pm	0.73	$&$	10.79	\pm	0.33	$&$	2.95	\pm	1.59	$&$	2.76	\pm	0.17	$&$	1.05	\pm	0.86	$	\\
67 &90110-02-01-02	&2004-08-15&$	5.18	\pm	0.11	$&$	0.51	\pm	0.21	$&$	10.97	\pm	0.34	$&$	2.36	\pm	0.17	$&$	3.39	\pm	0.35	$&$	1.68	\pm	0.21	$	\\
68 &92428-01-04-00	&2007-02-12&$	4.34	\pm	0.01	$&$	0.40	\pm	0.02	$&$	8.77	\pm	0.04	$&$	0.85	\pm	0.09	$&$	2.16	\pm	0.05	$&$	0.78	\pm	0.11	$	\\
69 &92428-01-04-01	&2007-02-12&$	4.21	\pm	0.02	$&$	0.31	\pm	0.04	$&$	8.47	\pm	0.05	$&$	0.82	\pm	0.13	$&$		 --		$&$		 --		$	\\
70 &92428-01-04-02	&2007-02-13&$	4.13	\pm	0.01	$&$	0.37	\pm	0.03	$&$	8.40	\pm	0.06	$&$	1.08	\pm	0.13	$&$	2.17	\pm	0.09	$&$	0.40	\pm	0.16	$	\\
71 &92428-01-04-03	&2007-02-13&$	5.05	\pm	0.02	$&$	0.50	\pm	0.04	$&$	10.05	\pm	0.06	$&$	0.95	\pm	0.17	$&$	2.54	\pm	0.06	$&$	0.38	\pm	0.13	$	\\
72 &92428-01-04-04	&2007-02-14&$	5.61	\pm	0.02	$&$	0.64	\pm	0.05	$&$	11.58	\pm	0.16	$&$	2.05	\pm	0.58	$&$	2.91	\pm	0.04	$&$	0.27	\pm	0.15	$	\\
73 &92035-01-02-01	&2007-02-02&$	0.26	\pm	0.01	$&$	0.04	\pm	0.01	$&$		--		$&$		--		$&$		--		$&$		--		$	\\
74 &92035-01-02-07	&2007-02-07&$	0.90	\pm	0.01	$&$	0.06	\pm	0.01	$&$		--		$&$		--		$&$		--		$&$		--		$	\\
75 &92035-01-02-08	&2007-02-06&$	0.55	\pm	0.01	$&$	0.06	\pm	0.03	$&$		--		$&$		--		$&$		--		$&$		--		$	\\
76 &92704-03-11-00	&2007-05-14&$	4.03	\pm	0.02	$&$	0.08	\pm	0.07	$&$		--		$&$		--		$&$		--		$&$		--		$	\\
77 &92704-03-11-01	&2007-05-15&$	3.25	\pm	0.09	$&$	0.32	\pm	0.15	$&$		--		$&$		--		$&$		--		$&$		--		$	\\
78 &92704-03-12-00	&2007-05-16&$	2.70	\pm	0.31	$&$	0.44	\pm	0.30	$&$		--		$&$		--		$&$		--		$&$		--		$	\\
79 &92704-03-12-01	&2007-05-17&$	2.70	\pm	0.05	$&$	0.37	\pm	0.28	$&$		--		$&$		--		$&$		--		$&$		--		$	\\
80 &92704-04-01-01	&2007-05-16&$	2.99	\pm	0.09	$&$	0.70	\pm	0.39	$&$		--		$&$		--		$&$		--		$&$		--		$	\\
81 &92704-04-01-02	&2007-05-16&$	2.75	\pm	0.12	$&$	0.43	\pm	0.39	$&$		--		$&$		--		$&$		--		$&$		--		$	\\
82 &92704-04-01-04	&2007-05-17&$	2.81	\pm	0.09	$&$	0.61	\pm	0.27	$&$		--		$&$		--		$&$		--		$&$		--		$	\\
83 &92704-04-01-05	&2007-05-17&$	2.93	\pm	0.22	$&$	1.02	\pm	0.69	$&$		--		$&$		--		$&$		--		$&$		--		$	\\
84 &95409-01-12-04	&2010-03-31&$	0.19	\pm	0.01	$&$	0.03	\pm	0.01	$&$		--		$&$		--		$&$		--		$&$		--		$	\\
85 &95409-01-13-00	&2010-04-03&$	0.27	\pm	0.01	$&$	0.04	\pm	0.01	$&$		--		$&$		--		$&$		--		$&$		--		$	\\
86 &95409-01-13-01	&2010-04-07&$	0.37	\pm	0.01	$&$	0.07	\pm	0.02	$&$		--		$&$		--		$&$		--		$&$		--		$	\\
87 &95409-01-13-02	&2010-04-05&$	0.32	\pm	0.01	$&$	0.05	\pm	0.02	$&$		--		$&$		--		$&$		--		$&$		--		$	\\
88 &95409-01-13-03	&2010-04-02&$	0.22	\pm	0.01	$&$	0.01	\pm	0.01	$&$		--		$&$		--		$&$		--		$&$		--		$	\\
89 &95409-01-13-04	&2010-04-04&$	0.29	\pm	0.01	$&$	0.03	\pm	0.01	$&$		--		$&$		--		$&$		--		$&$		--		$	\\
90 &95409-01-13-05	&2010-04-06&$	0.37	\pm	0.01	$&$	0.03	\pm	0.02	$&$		--		$&$		--		$&$		--		$&$		--		$	\\
91 &95409-01-13-06	&2010-04-08&$	0.45	\pm	0.01	$&$	0.04	\pm	0.02	$&$		--		$&$		--		$&$		--		$&$		--		$	\\
92 &95409-01-14-01	&2010-04-10&$	1.05	\pm	0.01	$&$	0.07	\pm	0.02	$&$		--		$&$		--		$&$		--		$&$		--		$	\\
93 &95409-01-14-02	&2010-04-11&$	2.49	\pm	0.01	$&$	0.16	\pm	0.02	$&$		--		$&$		--		$&$		--		$&$		--		$	\\
94 &95409-01-14-03	&2010-04-12&$	1.58	\pm	0.01	$&$	0.03	\pm	0.02	$&$      3.15   \pm	0.02	$&$	0.15	\pm     0.03	$&$		--		$&$		--		$	\\
95 &95409-01-15-01	&2010-04-17&$	5.66	\pm	0.02	$&$	0.28	\pm	0.09	$&$		--		$&$		--		$&$		--		$&$		--		$	\\
96 &95409-01-17-02	&2010-05-02&$	6.75	\pm	0.11	$&$	1.14	\pm	0.39	$&$		--		$&$		--		$&$		--		$&$		--		$	\\
97 &96409-01-06-01	&2011-02-06&$	4.56	\pm	0.18	$&$	1.51	\pm	0.63	$&$		--		$&$		--		$&$		--		$&$		--		$	\\
\label{table:obsdet}
\end{longtable}
\end{center}
\twocolumn
\end{document}